

\magnification=1200
\hsize = 5.5 true in
\hoffset = 0.5 true in
\baselineskip=20pt
\lineskiplimit=1pt
\lineskip=2pt plus .5pt
\parskip = 3pt
\font\tenrm=cmr10
\font\ninerm=cmr9

\def\today{\number\month /\number\day /\number\year}
\pageno=0
\headline={\rm \ifnum \pageno<2 LA-UR-93-4205 \hfil \today
		\else LA-UR-93-4205 \hfil
``Ion Pair Potentials-of-Mean-Force'' \hfil \today \fi}
\footline={\rm \ifnum \pageno<1 \hfil \else \hfil \folio \hfil \fi}

\topskip = 48pt
\parindent =  20pt
\vfil
\centerline{\bf Ion Pair Potentials-of-Mean-Force in Water}

\bigskip

\centerline{
	Lawrence R. Pratt, Gerhard Hummer, and Angel E. Garcia }
\centerline{Los Alamos National Laboratory, }
\centerline{Los Alamos, NM 87545}

\bigskip
\bigskip


\parindent =  20pt
\topskip = 10pt

\baselineskip=11pt
\ninerm
\centerline{ {\bf Abstract} }

	Motivated by the surprising dielectric model predictions of
alkali-halide ion pair potentials-of-mean-force in water due to
Rashin, we reanalyze the theoretical bases of that comparison.  We
discuss recent, pertinent molecular simulation and integral equation
results that have appeared for these systems.  We implement dielectric
model calculations to check the basic features of Rashin's
calculations.  We confirm that the characteristic structure of contact
and solvent-separated minima does appear in the dielectric model
results for the pair potentials-of-mean-force for oppositely charged
ions in water under physiological thermodynamic conditions.
Comparison of the dielectric model results with the most current
molecular level information indicates that the dielectric model does
not, however, provide an accurate description of these
potentials-of-mean-force.  Since literature results indicate that
dielectric models can be helpfully accurate on a coarse, or chemical
energy scale, we consider how they might be based more firmly on
molecular theory.  The objective is a parameterization better
controlled by molecular principles and thus better adapted to the
prediction of quantities of physical interest. Such a result might be
expected to describe better the thermal-level energy changes
associated with simple molecular rearrangements, {\it i.e.\/}, ion
pair potentials-of-mean-force.  We note that linear dielectric models
correspond to modelistic implementations of second-order thermodynamic
perturbation theory for the excess chemical potential of a
distinguished solute molecule.  Therefore, the molecular theory
corresponding to the dielectric models is second-order thermodynamic
perturbation theory for that excess chemical potential.  Examination
of the required formulae indicate that this corresponding molecular
theory should be quite amenable to computational implementation.  The
second-order, or fluctuation, term raises a technical computational
issue of treatment of long-ranged interactions similar to the one
which arises in calculation of the dielectric constant of the solvent.
Satisfactory calculation of that term will require additional
theoretical consideration of those issues.  It is contended that the
most important step for further development of dielectric models would
be a separate assessment of the first-order perturbative term
(equivalently the {\it potential at zero charge} ) which vanishes in
the dielectric models but is generally nonzero.  Parameterization of
radii and molecular volumes should then be based of the second-order
perturbative term alone.  Illustrative initial calculations are
presented and discussed.

\tenrm
\baselineskip=20pt

\vfil
\eject

\bigskip
\noindent
{\bf Introduction}

	Potentials describing the average forces on a set of solution
constituents can be taken as primitive ingredients in molecular
theories of solutions [1]. Biology typically involves aqueous
solutions and much of the physics of such solutions requires the
treatment of strong electrolytes.  Thus, the potentials of the average
forces among ionic species in aqueous solution are of special
relevance to theories of biomolecular structure and dynamics.

	Ion pairs form the simplest set of ionic solution constituents for
which the potentials of average forces are non-simple.  It is quite
natural, therefore, that much theoretical and computational effort has
been expended in determining those quantities from a molecular basis.
Molecularly detailed calculations have been pursued deliberately for
the past 15 years [2] and we are nearing the stage of having reliable
molecular results for the simplest ion pairs under an important but
limited range of solution conditions.

	Because of the difficulty of the molecular calculations a simple,
physical model of those ionic pair potentials-of-mean-force which
reliably captured the important physical features of the known
molecular data would be extremely valuable.  From this perspective the
results of Rashin's application [3] of a dielectric model to the
prediction of the potentials of the mean forces between alkali-halide
ion pairs in water are especially enticing.  Rashin calculated those
quantities for a dielectric model and compared the results with
integral equation theories for the same quantities.  Quite
surprisingly, the compared quantities were qualitatively very similar.
In particular, the oscillatory features that are often seen in the
potentials of the mean forces between atoms or ions in dense liquids
appeared in a dielectric model also.  The two most significant
potential wells were located in about the right place on the basis of
a dielectric model and the relative depths of those wells seemed also
to be qualitatively correct in the model.  A more quantitative
analysis was not given.

	This paper develops a more current and quantitative analysis.
Important new molecular results on ion pair potentials-of-mean-force
in water have appeared since Rashin's work.  We discuss those recent
molecular simulation information which are pertinent to the issue but
do not attempt a conventional review.  We then implement a dielectric
model for the physical circumstances studied by Rashin in order to
confirm the essential correctness of those results.  There are some
technical differences between our model calculation and Rashin's that
will check whether the important features of Rashin's results are
dependent on minor changes in the model.  At that stage we compare the
molecular results with those of the model and draw some revised
conclusions.  We consider how the parameterization of a dielectric
model, in particular the size and shape of the molecular cavity, might
be modified to bring the predictions of the model more closely into
agreement with the molecular results.  Finally, we attempt to identify
what the qualitative success of a dielectric dielectric model teaches
us about how to do more effective molecular calculations and how that
lesson helps us in achieving a physically correct parameterization of
the model.

\bigskip \noindent {\bf Molecular Results on Potentials-of-Mean-Force
between Simple Ion Pairs in Water}

\noindent {\it Motivation.} The explicit molecular modeling of the
solvent water in the computation of electrolyte properties gives
access to such important quantities as the hydration structure of ions
and the dielectric screening of inter-ionic interactions as it depends
on solution conditions. In the primitive models the ions are described
as rigid spheres that, outside of overlap, interact with the classic
long-ranged $1/\varepsilon r$ electrostatic interaction screened by
the macroscopically measured solvent dielectric constant.  In more
flexible primitive model descriptions, solvent properties are still
simple input parameters, {\it e.g.}, water molecules might be viewed
as forming an impenetrable first hydration layer around the ions
whereas bulk water might be modeled as a homogeneous medium with high
dielectric constant independent of the actual ionic concentration. But
the effects of the molecular ``granularity'' of the solvent on the
potentials-of-mean-force (PMF) between ions require more painstaking
approaches. Such effects are expected to be particularly important at
short distances. The expected oscillatory behavior of the PMF's is due
not only to packing of hydrated ions --- perhaps well described in
primitive models --- but also due to packing of water molecules that
are not explicitly involved in the primitive models.

\smallskip \noindent {\it Difficulties in the calculation of
structural and thermodynamic properties.} Compared to the case of
simple liquids and primitive model descriptions of electrolytes,
reliable theoretical tools for calculating structural and
thermodynamic properties of electrolyte models with water included as
a molecule are underdeveloped. The well-established methods for atomic
liquids are not readily applicable for relatively complex molecular
systems. The reference interaction site model (RISM) integral
equations, developed to calculate the structure of molecular fluids at
an atom-atom correlation level, perform reasonably well for systems
with purely repulsive forces. But in the case of aqueous-electrolyte
solutions additional difficulties are encountered.

	Similarly, the applicability of the computer simulation methodology
is limited. One problem that arises is that in low and intermediate
concentration electrolyte solutions the number of water molecules
exceeds the number of ions by far. This requires the study of very
large size systems. The deep electrostatic interaction energies
associated with ionic and hydrogen-bonding interactions in addition to
the liquid state disorder means that long simulations are necessary to
get sufficient statistical accuracy. Moreover, the strong Coulomb
interactions also require particular attention because of their long
range. However, in primitive model descriptions the implicit high
dielectric screening that is appropriate allows the use of less
involved methods to compute the Coulomb interactions.

\smallskip \noindent {\it XRISM integral equation studies of {\it
Na}-{\it Cl}-{\it H}$_2${\it O} electrolytes.} The formulation of
Ornstein-Zernike (OZ) type integral equation schemes at the level of
atom-atom correlation functions of molecules has been extensively
analysed. For reviews see Chandler [4], Cummings and Stell [5], and
Monson and Morriss [6]. The reference interaction site model (RISM) is
the prototype of these approaches. It adopts an atom-atom
Ornstein-Zernike (RISM-OZ) equation that relates total atom-atom
correlations $h_{ij}(r)$, atom-atom {\it direct\/} correlations
$c_{ij}(r)$, and intra-molecular correlations $\omega_{ij}(r)$. This
equation defines the $c_{ij}(r)$ in terms of the other quantities that
are, in principle, measurable. A second relation, a {\it closure
relation}, is essential to obtaining equations that might be the basis
of a numerical solution effort. The discovery of physically motivated,
mathematically controlled closures is a serious problem.
Well-established closures in the context of Ornstein-Zernike equation
theories of atomic liquids have been transfered by analogy to the
atom-atom level treatment of molecular liquids. Analogues of the mean
spherical approximation, the Percus-Yevick theory, and hypernetted
chain (HNC) closure are available. However, applied to the atom-atom
level description of molecular liquids, the theoretical justification
of all these analogues is more delicate than is the case for atomic
liquids. For example, it has been shown that the RISM theory, when
viewed from the perspective of graphical cluster expansions series,
includes ``improper'' diagrams. For a discussion, see Chandler, {\it
et al.\/} [7], and Monson and Morriss [6]. Therefore, the search for
physically motivated, mathematically controlled closures of the
RISM-OZ equation is of considerable importance [8].

In the application of the RISM method to polar and ionic systems,
additional difficulties arise. For atomic liquids the assumption that
the direct correlation functions asymptotically vary like the
potential, $c(r)\rightarrow -\beta u(r)$ for $r\rightarrow\infty$ is
well-justified.  However, transfered to the atom-atom level for
molecular liquids on the basis of the RISM-OZ equation,
$c_{ij}(r)\rightarrow -\beta u_{ij}(r)$ for $r\rightarrow\infty$
results in a seriously wrong dielectric behavior. The dielectric
constant assumes an ideal gas value which is too low. Again, the
problem may be reduced to finding an appropriate closure to which the
asymptotic behavior of $c_{ij}(r)$ is intimately connected.

For special cases such as polar, diatomic molecules, the asymptotic
behavior of $c_{ij}(r)$ has been studied carefully [5]. For general
molecular structures such an analysis is not currently available. A
brute-force method to ``improve'' the dielectric behavior consists in
modifying the $k^2$ term of the small-$k$ expansion of the atom-atom
total correlation functions in Fourier space [9,10] to which the
dielectric constant is related. It is clear that this approach
contains some arbitrariness and in addition requires {\it a priori}
knowledge about the dielectric behavior. Moreover, these modifications
affect the $r\rightarrow\infty$ behavior of the atom-atom direct
correlation functions requiring corrections, {\it e.g.}, by simply
neglecting the additional terms.

In view of these difficult theoretical issues, the results of
practical calculations are enlightening. The RISM method in
conjunction with the HNC closure --- the {\it X}RISM approximation
[11] --- has been applied to studying aqueous solutions of ions
employing interaction site descriptions of water molecules. Pettitt
and Rossky [12] analysed the structure of ion-water and ion-ion
correlations at infinite ionic dilution, where the XRISM equations
decouple. As a direct manifestation of the wrong dielectric behavior,
the infinite dilution ion-ion PMF's behave as $1/\varepsilon_{\rm
XRISM}r$ for $r\rightarrow\infty$, with $\varepsilon_{\rm XRISM}$
being approximately 18 instead of the water value of $\varepsilon
\approx 80$. By subtraction of the $1/\varepsilon_{\rm XRISM}r$ terms
and addition of $1/\varepsilon r$ terms, Pettitt and Rossky corrected
this behavior approximately. In the case of {\it Na}$^+$-{\it Cl}$^-$
solutions, the authors observed some remarkable features in the
ion-ion PMF's. The {\it Na}$^+\cdots${\it Cl}$^-$ PMF showed a deep
contact minimum and a deep, broad solvent-separated minimum. Most
interestingly, also found was a very deep contact minimum in the {\it
Cl}$^-\cdots${\it Cl}$^-$ PMF at about 0.35~nm distance. It is about
$-1.5kT$ deep and followed by a barrier with a PMF value of more than
$2.5kT$. The minimum in the {\it Cl}$^-\cdots${\it Cl}$^-$ PMF
reflects a strong tendency of {\it Cl}$^-$ ions to form contact pairs.
This astonishing XRISM result has sparked considerable debate. For a
discussion see Friedman, {\it et al.\/} [13]. However, recent computer
simulation studies [14] have clearly established that this surprising
feature of paired anions in the XRISM theory is not correct.

Recently, the XRISM equations for aqueous {\it Na}$^+$-{\it Cl}$^-$
solutions were also solved at finite concentration [9], {\it i.e.},
for ionic concentrations above 0.2~mol/l. (The concentration regime
between about $10^{-6}$ and 0.2~mol/l was observed to be inaccessible
to numerical solution. This likely reflects the unphysically weak
dielectric screening of the XRISM approximation.) The interaction
potentials of Pettitt and Rossky were used, except for a slightly
different water model. In view of the experimental decrease of the
dielectric constant with increasing salt concentrations, the
deficiencies of XRISM due to errors in the description of dielectric
screening might become less important under high salt conditions.
Indeed, quite reasonable results for the strucure were obtained.
However, XRISM predicts strongly paired {\it Cl}$^-$ ions under high
salt conditions also.

Perkyns and Pettitt studied an XRISM model that was modified in its
dielectric behavior [10]. Utilizing a result for the total atom-atom
correlation functions $h_{ij}$ by H{\o}ye and Stell [15], they
modified the small-$k$ behavior of the Fourier transformed solvent
atom-atom correlation functions by adding a function
$\tilde{\chi}_{ij}(k)$ decaying as $\exp(-\alpha k^2)$ with $\alpha$
an adjustable parameter.  Any modification in the total atom-atom
correlation functions has its counterpart $\tilde{b}_{ij}(k)$ in the
atom-atom direct correlation functions via the RISM-OZ equations. In
particular, small-$k$ modifications of the correlation functions
affect the long range behavior of the atom-atom direct correlation
functions. To overcome these problems, the authors simply modify the
HNC closure relation so that the additional terms are exactly
cancelled. In Ref.~10(a), results at 1~mol/l concentration for the
{\it Na}$^+\cdots${\it Cl}$^-$ pair correlation function were
described using the same model as in Ref.~9.  The dielectric constant
implicit in their calculation was changed from approximately 18 (the
XRISM value) to 78 and they observed a marked diminution of the height
of the contact peak.  In Ref. 10(b), the authors discuss results for a
different interaction site model of aqueous {\it Na}$^+$-{\it Cl}$^-$
solutions, again at 1~mol/l concentration. For this model (with
$\varepsilon$ modified to 78.54) they find a very deep contact minimum
in the {\it Na}$^+\cdots${\it Cl}$^-$ PMF of approximately $-2.6kT$.
The {\it Cl}$^-\cdots${\it Cl}$^-$ PMF shows a minimum of about
$-0.7kT$ at 0.4~nm distance.  Unfortunately, no comparable results for
the {\it Cl}$^-\cdots${\it Cl}$^-$ pair correlation had been given in
Ref. 10(a).  It would be interesting to see the corresponding results
of the model of Ref. 10(a) for comparison, in order to disentangle
effects of different models and of limitations of the RISM
formulation.

Another limitation in the RISM formulation is that intramolecular
structure is described only approximately. Thus, some elements of
structural consistency between between different correlation functions
are not obeyed. This unsatisfactory effect is evident in the positions
of peaks in the correlation functions of hydrogen with negatively
charged sites (water-oxygen and {\it Cl}$^-$ ions in aqueous {\it
Na}$^+$-{\it Cl}$^-$ solutions) and the corresponding results for
water-oxygen. For example, the distance between the first peaks of the
{\it H}$\cdots${\it Cl}$^-$ and {\it O}$\cdots${\it Cl}$^-$
correlation function significantly exceed the length of the rigid {\it
O}-{\it H} bond. This reflects the difficulty of the RISM scheme with
presently used closure relations to correctly transfer the repulsive
interaction of the oxygen site with a negatively charged site to the
strongly attracted hydrogen site on a rigid water molecule. In the
interaction site water models studied extensively by computer
simulation, the hydrogen site typically has only a weak or no
repulsive shell of its own, but is ``protected'' inside the van der
Waals radius of the oxygen. Such bond constraints violations are also
observed in a dielectrically modified XRISM formulation; {\it cf.\/}
Fig.~3 of Ref. 10(b).

In view of the discussed short-comings, the XRISM methodology for the
calculation of structural and thermodynamic properties of strongly
associating, hydrogen bonding liquids such as aqueous electrolytes, is
of limited utility. Particular problems are ({\it i}) the lack of
physically motivated, mathematically controlled closure relations,
({\it ii}) the deficiencies in the dielectric screening behavior, so
far attacked only by {\it ad hoc} modifications of the theory; and
({\it iii}) the violation of bond constraints in hydrogen bonds, that
are essential for describing the hydration of anions and the formation
of hydrogen bonds between water molecules correctly.  Aside from these
arguments, the restricted reliability of presently available XRISM
integral equation schemes in describing aqueous electrolytes becomes
evident by comparison with extensive computer simulations of two high
concentration {\it Na}$^+$-{\it Cl}$^-$-water electrolytes [14]. This
comparison clearly shows that XRISM results for ion-ion PMF's should
be considered with caution.  This now recognized point is clearly of
relevance to our reexamination of the comparison presented by Rashin
several years earlier.

However, the further development of integral equation schemes
describing complex molecular fluids, particularly aqueous solutions,
is very important. Many important quantities like solute-solute
correlations are not readily accessible with experimental techniques.
Computer simulations of aqueous phases are usually very costly with
respect to CPU time; and the calculation of thermodynamic quantities
is usually subject to very large statistical uncertainties in the case
of electrolytes. Integral equation methods such as an improved RISM
scheme would open the biologically important low and intermediate
ionic concentration regime to the theoretical analysis.

\smallskip \noindent {\it Computer simulation studies of {\it Na}-{\it
Cl}-{\it H}$_2${\it O} electrolytes.} In a series of computer
simulations, two high concentration aqueous sodium-chloride
elect\-ro\-lyte systems were analysed, one at room temperature
[14(b)], the other at 823~K [14(a)].  In both cases, the SPC model of
water was used [16], in conjunction with the parameters of Pettitt and
Rossky for the ion-water and ion-ion interaction [12]. The ionic
concentration in the room temperature system was 5~mol/l. The high
temperature system consisted of 20 mass percent {\it Na}$^+$-{\it
Cl}$^-$ and had a total mass density of 0.867~g/cm$^3$, corresponding
to an experimentally observed pressure of 2500~bar.  We discuss these
results particularly rather than review simulation calculations on
these problems more broadly for several reasons.  These results have
appeared very recently, they focused on comparative testing of
alternative treatments of long-ranged interactions, and XRISM
calculations were performed on the same interaction models.  The
results should depend on the interaction models used.  The sensitivity
of the dependence of the {\it Cl}$^- \cdots${\it Cl}$^-$ and {\it
Na}$^+ \cdots${\it Na}$^+$ pair has been observed by Dang, {\it et
al.\/} [17].

The results described in the following were obtained with constant
energy molecular dynamics simulations of 500 particle systems (416 and
432 water molecules in the room temperature and high temperature
simulation, respectively). The standard Verlet integration method was
used along with constraints for the rigid water model [18]. A time
step of 0.005~ps was employed with the hydrogen masses set to 2~g/mol
(heavy water). The electrostatic interactions were treated with Ewald
summation [18] using $2\times 510$ ${\bf k}$-vectors and a real space
damping factor of $5.6/L$.

Fig.~(1) shows the ion-ion PMF's, calculated as $-kT\log g_{ij}(r)$
for these simulations at finite concentrations of salt.  These results
are in good agreement with the calculations of Gu\`{a}dria, {\it et
al.\/}, [19] who focused explicitly on infinite dilution conditions by
treating only one ion-pair in their simulation volume.  Those
calculations used the same model for interactions among water (SPC)
with the exception that they permitted some internal flexibility of
the water molecule structure.  We also note that Hummer, {\it et
al.\/}, [20] have used just the same models and methods to study the
concentration dependence of the distribution functions $g_{ij}(r)$ in
the range 1 to 5 molar. Our discussions here are consistent with those
recent results.  As expected, drastic differences between the two
thermodynamic states are observed. The high temperature PMF's are
generally less structured than the corresponding room temperature
curves. This can be understood as a consequence of the comparably weak
hydration shell of ions at 823~K, which can easily be disrupted by
thermal excitations.

The high temperature {\it Na}$^+\cdots${\it Na}$^+$ PMF shows very
shallow minima at about 0.5 and 0.7~nm distance, whereas the room
temperature curve has minima at 0.37~nm ($0.5kT$) and at 0.6~nm
distance ($-0.2kT$), which are clearly separated by a barrier
($1.2kT$). Interestingly, in the repulsive region (for distances below
0.37~nm) the two curves are very similar, a behavior that is not found
in the case of the {\it Na}$^+\cdots${\it Cl}$^-$ and {\it
Cl}$^-\cdots${\it Cl}$^-$ PMF's. The well-separated minima in the room
temperature {\it Na}$^+\cdots${\it Na}$^+$ PMF can be interpreted as
{\it contact} and {\it solvent-separated} states. Pairs of positively
charged {\it Na}$^+$ ions at small distance can be formed at room
temperature, with negatively charged water oxygens mediating the
repulsion. However, the {\it contact} minimum in the {\it
Na}$^+\cdots${\it Na}$^+$ PMF is only weakly populated with a PMF
value of $0.5kT$.

We also observe qualitative differences between the PMF's for {\it
Na}$^+$ $\cdots$ {\it Cl}$^-$ pairs at the two thermodynamic states.
At 823~K the contact minimum at 0.26~nm distance is highly populated
with a PMF value of $-2.35kT$, whereas at room temperature the
corresponding PMF value is only $0.56kT$. On the other hand, the room
temperature solvent-separated minimum at about 0.5~nm distance is very
broad and deep ($-1.0kT$), in contrast to the shallow minimum in the
823~K curve at 0.53~nm distance. The tendency to form well-separated
{\it Na}$^+\cdots${\it Cl}$^-$ ion pairs at room temperature reflects
a high stability of the first hydration shell of the ions. Even large
Coulomb energy gains of bringing oppositely charged ions together do
not suffice to form a considerable amount of {\it Na}$^+\cdots${\it
Cl}$^-$ pairs in contact.

The shapes of the {\it Cl}$^-\cdots${\it Cl}$^-$ PMF's at room
temperature and at 823~K are very similar, with the room temperature
curve shifted about 0.1~nm towards larger distances. Both curves show
a shallow first minimum (0.54~nm and $-0.37kT$ at room temperature but
0.45~nm and $-0.25kT$ at 823~K) followed by maxima of $0.31kT$ (room
temperature) and $0.16kT$ (823~K).

A formation of {\it Cl}$^-$ pairs in contact is not observed at 823~K,
in contradiction to the XRISM prediction [14(a)]. At 823~K, XRISM
shows a deep minimum in the {\it Cl}$^-\cdots${\it Cl}$^-$ PMF at
approximately 0.36~nm distance with a PMF value of $-0.44kT$. Even
stronger disagreement between the XRISM result and the corresponding
simulation data is observed at room temperature. The simulation shows
{\it Cl}$^-$ ions always well separated. However, the corresponding
XRISM curve exhibits a very distinct minimum of almost $-1.5kT$ at
0.34~nm distance [9].

Summarizing the simulation results for ion-ion PMF's at high ionic
concentration, at 823~K thermal excitations are able to disrupt the
ionic hydration shell. The contact minimum in the {\it
Na}$^+\cdots${\it Cl}$^-$ PMF is well populated. At room temperature
{\it Na}$^+$ and {\it Cl}$^-$ ions tend to remain independently
solvated. They hardly come closer than the diameter of a hydrated ion.
Correspondingly, the deepest minima of the PMF's are found at
distances of 0.6~nm ({\it Na}$^+\cdots${\it Na}$^+$), 0.5~nm ({\it
Na}$^+\cdots${\it Cl}$^-$), and 0.54~nm ({\it Cl}$^-\cdots${\it
Cl}$^-$). Evidently, the correct description of this behavior is a
serious challenge for any theoretical model, since the ion-ion PMF's
are the result of a thermal balance of competing strong interactions.
All water-water, water-ion, and ion-ion interactions give relevant
contributions. For instance, severe problems are expected in the case
of the {\it Na}$^+\cdots${\it Cl}$^-$ PMF, which shows a preference
for the solvent-separated state, in spite of the huge Coulomb
attraction of the two oppositely charged ions. But also the accurate
calculation of correlations of like charged ions is far from trivial,
as can be seen from the qualitatively and quantitatively wrong results
of the XRISM formulation for the {\it Cl}$^-\cdots${\it Cl}$^-$ PMF.

\bigskip \noindent {\bf Continuum Model Prediction of the $Na^+ \cdots
Cl^-$ Potential-of-Mean-Force in Water}

Dielectric models can capture gross electrostatic effects with a crude
veracity.  They can be physical models in the sense that a molecular
system, albeit highly idealized, can be identified that would exhibit
just the properties ascribed to a dielectric model.  In this quality
the dielectric models stand in sharp contrast with available integral
equation theories. Those integral equation theories offer the
potential of fuller incorporation of molecular detail as the price
paid for a more tortuous connection to simple physical concepts.  They
are are also more complicated by their utilization of additional
molecular detail.  Much of the detail required to implement the
molecular calculations is absent in the dielectric models.  Thus,
Rashin's observation that a dielectric model potential-of-mean-force
for alkali-halide ion pairs in water is in good qualitative agreement
with the available XRISM results was surprising.

	In particular, features of the molecular results which would almost
universally be interpreted as due to molecular packing were also
predicted by the dielectric model.  Since dielectric models depend on
empirical parameterization of molecular surfaces, the observed
agreement might not seem compelling by itself.  However, Rashin's
surprising observation raises the possiblity that our understanding of
these potentials-of-mean-force can be made more physical and
simplified.  A confirmation of the principal features of Rashin's
result would then focus attention on parameterization and
implementation of dielectric models to make the best use of available
molecular scale information.  Since realistic molecular models are
different from the dielectric ones, the dielectric models should be
expected to fail when aggressively pushed into more molecular regimes.
Even in such cases it can be reasonably hoped that dielectric models
would provide a helpfully accurate initial approximation to which
molecular scale refinements could be added.

	Therefore, we have performed calculations similar to Rashin's for the
$Na^+ \cdots Cl^-$ ion pair in water.  We solved $$\nabla \bullet
\varepsilon ({\bf r})\nabla \Phi ({\bf r}) = - 4\pi \rho _f({\bf r})
,\eqno{(1)}$$ where $\varepsilon({\bf r})$ is one inside, and the
solvent dielectric constant $\varepsilon$ outside, the bipolar volume
depicted in Fig.~(2). The free charge $\rho_f({\bf r})$ describes
$\delta$-function sources of strength $+e$, ($-e$), at the $Na^+$,
($Cl^-$) nucleus.  The radii of the solvent exclusion spheres centered
on these nuclei were taken from Rashin and Honig [21]: $R_{Na^+} =
0.1680 $ nm \ and $R_{Cl^+} = 0.1937$ nm. The value of the dielectric
constant outside the molecular volume was $\varepsilon = 77.4$, the
measured value of the dielectric constant for water at its triple
point.  The difference between the model treated here and by Rashin is
that in Rashin's case the atom centered spheres are joined by a
toroidal surface fragment when the shortest distance between the
spherical surfaces is smaller than a van der Waals diameter of a water
molecule.

	The method of numerical solution was essentially a boundary element
approach [22-28] based upon an integral equation formulation of
Eq.~(1): $$\varepsilon({\bf r})G({\bf r},{\bf r}')= G_0({\bf r},{\bf
r}') +\int\limits_V { {\nabla ''G_0({\bf r},{\bf r}'')} \bullet \left[
{{{\nabla ''\varepsilon ({\bf r}'')} \over {4\pi \varepsilon ({\bf
r}'')}}} \right] {\varepsilon ({\bf r}'')G({\bf r}'',{\bf r}')}
d^3r''} . \eqno{(2)}$$ Here $G({\bf r},{\bf r}')$ is the Green
function for the problem, {\it i.e.\/}, the electric potential at
${\bf r}$ due to a unit source at ${\bf r}'$. $G_0({\bf r},{\bf r}')$
is the Green function for the $\varepsilon({\bf r})=1$ case.  This
general equation is correct either for zero boundary data on a surface
everywhere distant or for periodic boundary conditions with a system
of volume $V$.  The article by Yoon and Lenhoff [26] can be consulted
for a statement of the advantages of boundary approaches over
alternatives which discretize the three dimensional domain.  The
present approach was chosen because it is simple to see how the
integral equation can be solved by Monte Carlo methods also and that
possibility holds some potential for reliable application to much
larger systems.  In fact, achieving numerical convergence for the
solvation free energy on a thermal energy scale is not trivial.  The
variations of interest here are typically less than 1\% of the total
solvation free energy of the complex.  Similarly accurate results for
very much larger systems will provide a computational challenge.  The
details of the numerical methods used by Rashin are clearly quite
different from those used here but those issues are not important for
the present discussion.  Once this equation is solved the free energy
of solvation (excess chemical potential), $\Delta \mu(r)$ which
depends on the separation of the sphere centers, and is obtained as:
$$\Delta \mu(r) =({1 \over 2}) \sum_{i,j} q_i q_j [G({\bf r}_i,{\bf
r}_j)-G_0({\bf r}_i,{\bf r}_j)] \eqno{(3)}$$ where the $q_i$ are the
charges associated with $\rho_f({\bf r})$.  The
potential-of-mean-force, $w(r)$, is then composed of this solvation
free energy {\it plus\/} the bare ion-ion potential energy of
interaction.  For this latter quantity we used the same function as
Rashin did.

	The results are shown in Fig.~(2).  Studies of numerical convergence
with respect to the resolution of the sampling of the molecular
surface leads us to the conclusion that the curve shown there is
accurate to within about $0.3 kT$.  However, all the features of the
model known {\it a priori\/} are correctly described in this numerical
solution and all qualitative features of the curve are insensitive to
further enhancement of that resolution.  In particular, the positions
and depths of the two wells and the position and height of the barrier
are not appreciably changed by changes in the surface density of the
boundary sampling by a factor four.

This model result is in good qualitative agreement with that of Rashin
except for the height of the free energy barrier between the contact
and the outer minima.  The barrier is much higher in Rashin's case;
that is surely due to the slight differences in definition of the
volume of the ion-pair, particularly the `neck' between the two atom
centered spheres utilized by Rashin but not by us.  The molecular
volume becomes disconnected within the region of the barrier.  The
position and depth of the inner-most free energy well is in very
accurate agreement with Rashin's result. For separations near the
contact minimum the neck feature of Rashin's molecular surface is
graphically insignificant. The present comparison establishes that it
is also insignificant for prediction of the free energy of the contact
pair.

	It is natural to think that the repulsive portion of this
potential-of-mean-force is due principally to van der Waals overlap of
the ions.  This is not the case for the dielectric model result.
Fig.~(3) shows the separate contributions, solvation free energy and
inter-ionic potential energy, to the potential-of-mean-force.  It is
clear that the electrostatic solvation, solely, pulls the ions apart
for distances larger than about $r \approx 0.23$ nm.  The outer
minimum probably does not have a general structural explanation.  We
have found that this minimum can disappear if either the dielectric
constant or the radii assumed for the ionic cavities are changed
substantially.  However, the ions lose stabilizing solvation free
energy when they are brought together.  The outer minimum is a result
of the balancing of solvation stabilization of the separate ions with
the bare inter-ionic attraction.  This argument clearly does not apply
to ions of the same formal charge.  In fact, the dielectric model
results for those potentials-of-mean-force are qualitatively unlike
the results of Fig.~(1).  For example, the $Cl^- \cdots Cl^-$ pair
potential-of-mean-force predicted by the dielectric model is entirely
positive. It is strongly repulsive but only after overlap of the
separate ionic volumes near $r=0.39$ nm.  At larger distances, that
model result is more repulsive than the molecular result of Fig.~(1)
but the model result conforms accurately to the expected asymptotic
form of $q^2/ \varepsilon r$.

	Although the detail obtained from the prediction of the dielectric
model is surprising, it is not in accurate agreement with the
molecular results discussed in the previous section.  Compare Fig.~(2)
with the middle panel of Fig.~(1).  The contact well is much too deep
relative to the asymptote in the dielectric model.  One way to think
about the disagreement is the following: The radii which are adopted
here for definition of the molecular volume of the ion pair are ones
which produce reasonable values for the solvation free energies of the
separated ions.  But if one set of radii must be used for all
separations $r$ then the way to make the contact well less deep
relative to the asymptote is to lower the asymptote by increasing the
stabilization of the widely separated ions, {\it i.e.\/}, to {\it
decrease\/} the radii used.  But that doesn't seem reasonable.
Alternatively, we could adjust the radii for each ion pair separation
--- increasing the radii as the inter-ionic separation decreases ---
to get the correct solvation free energy. Much of the utility of the
model would then be lost.  These comments highlight the fact that the
ionic radii are non-trivial parameters of the dielectric model and
this parameterization is best considered from the perspective of
molecular theory.

In summary, we conclude that the model which produces the result of
Fig.~(2) fails to reproduce the molecular $Na^+ \cdots Cl^-$
potential-of-mean-force in water, in the absence of further
adjustments or theoretical refinements.  The most significant failure
is in the depth of the contact minimum, or equivalently the free
energy of the contact pair relative to the separated ions.  Since the
present calculation and Rashin's accurately agree in the neighborhood
of that contact minimum this most significant error is not due to
differences between the two calculations in definition of the
molecular surface.  The next section considers some further steps for
the molecular theory which offer some potential for improvement of the
present dielectric model approach.

\bigskip \noindent {\bf What Can We Learn from Comparison of Molecular
with Continuum Model Results?}

	On the basis of numerous calculations [29,30] in addition to the
results presented here, we conclude that dielectric models provide a
physical description of the solvation of charged and polar species in
solution but with great crudity.  Here we address the question of what
we should learn from these comparisons and how we can take advantage
of those lessons to obtain more reliable and efficient theoretical
predictions.

	As a focus for this discussion we use the prediction of the
dielectric model for the solvation free energy of a spherical ion: $$
\Delta \mu = - ( {q^2 \over 2 R} ) \times ( {\varepsilon -1 \over
\varepsilon } ) . \eqno{(4)}$$ This is the classic Born model.  Here
the ion has charge $q$ and a radius value of $R$ is adopted.  $q$ and
$\varepsilon$ are parameters that are scarely ever subject to
adjustment.  This is not true of $R$, however.  A good value for this
parameter is not immediately evident.  This can be recognized by
considering that the solvation free energy of the ion can be
determined exactly if the medium is a weakly polar, dilute gas.  In
that case, $R$ is the distance within which no polarization is ever
observed, {\it i.e.\/}, a distance of closest approach.  But this is
not the value of $R$ which works for the ion in liquid water.
Similarly, if we consider solvation properties of a particular solute
obtained for two different solvents with equal dielectric constants,
say methanol and acetonitrile, we must expect that the best molecular
volumes will be somewhat different in the two cases.  This point was
made already by Rashin and Honig [21]. But it is important to
emphasize the obvious and general conclusion that good values of $R$
depend on the solution conditions and are not a property of the ion
alone.  A corollary of this conclusion is that evaluation of
thermodynamic derivatives of the Born model $\Delta \mu$ should
account also for variations of $R$ with solution conditions.  That $R$
must be considered temperature dependent for the calculation of
solvation enthalpies has been pointed-out recently by Roux, {\it et
al.\/}.  [31]

	However once a reasonable value of $R$ is established experience
suggests that the dielectric model can be helpfully accurate.  The
calculation of Jayaram, {\it et al.\/} [32], gives contemporary
support to this idea. The conclusions we identify on this basis are:
(i) that the dependence of the solvation free energy on $q^2$ is
satisfactory; (ii) that the dependence on the solvent dielectric
constant $\varepsilon$ is a simple incorporation of rather
sophisticated information about the solvent and a better theory should
still take advantage of exactly known properties of the solvent; and
(iii) that the most pressing requirement for molecular theory is to
treat the molecular scale structure which is historically subsumed in
the parameter $R$.  The first two points here lead us directly to
second order perturbation theory for the excess chemical potential.
The required formula is $$ \Delta \mu \approx \Delta \mu _0 + \left<
\sum_j \varphi({\bf j}) \right> _0 -{\beta \over 2} \left< {\left(
\sum_j \varphi ({\bf j}) - \left< \sum_k \varphi ({\bf k}) \right> _0
\right)}^2 \right> _0 .  \eqno{(5)}$$ This approximation has been
discussed by Levy, {\it et al\/} [33].  It accounts only for the
effects of electrostatic perturbations $\varphi({\bf j})$ in the
interactions between an identified solute molecule, {\it e.g.\/}, the
{\it 0th\/} solute molecule, and the {\it jth\/} solvent molecule;
$\Delta \mu_0$ is the excess chemical potential of the designated
molecule in the absence of such interactions.  The brackets $\left<
\cdots \right>_0$ indicate the average value without the
perturbations.  It must be remembered that we have some flexibility in
definition of $\varphi({\bf j})$ at short range and that such a
definition can substantially affect the convergence of a perturbative
theory.  This idea underlies the WCA approach to the theory of simple
liquids.

	This formula has the properties sought.  First, the computed free
energy will be a general quadratic function of partial charges if
those quantites are used to describe the electrical charge
distribution of the solute.  The perturbative approach will supply
contributions independent of, and linear in, those charges; those
contributions are generally present though absent in the dielectric
model. Second, the correlation functions of the solvent, including
those correlations which establish the dielectric constant, are
exploited directly on the basis of molecular calculations or
information not further modeled or approximated.  Finally, the
solute-solvent molecular structures are treated directly on a
molecular level.

	Because the first-order term of Eq.~(5) stands out as a difference
between the molecular and the dielectric approach, the importance of
that contribution deserves emphasis.  $\Delta \mu$ will include a
contribution linear in $q$ which is associated with the electrostatic
potential of the phase and is of interest to questions of the
electrostatic potential difference between coexisting phases [34].
That contribution is not directly relevant to the local solvation
contributions of interest for this discussion.  Here we lump that
contribution together with $\Delta \mu_0$.  Beyond that contribution,
the first-order perturbation term is still non-trivial and generally
non-vanishing.  To the extent that this term is significant in
particular cases, merely adjusting radii to bring dielectric and
molecular calculations into agreement is physically misleading.  The
first-order contribution in the molecular approximation serves to
discriminate between positive and negative ions of the same physical
size; adjusting sizes to accomplish that discrimination in such cases
would confuse the origin of that effect.  It is possible that the
first-order contribution is principally involved in the disagreement
noted by Wilson and Pohorille [35] between dielectric model and
simulation results for the interaction of monovalent ions with the
water liquid-vapor interface.  The difference between positive and
negative ions of similar size seemed to be basic to their
observations.

	Although it is quite possible for the first-order term to contribute
significantly to the discrepancy between the dielectric model and
molecular results for the $Na^+ \cdots Cl^-$ ion pair
potential-of-mean-force, that first-order term is even more likely to
make an important contribution for the like-ion pair cases $Na^+
\cdots Na^+$ and $Cl^- \cdots Cl^-$.  For those cases the current
predictions of dielectric models are less satisfactory yet than for
the unlike-ion pair $Na^+ \cdots Cl^-$.  It would seem natural to
ascertain the contribution of the linear term before considering
effects which might be ascribed to perturbative contributions beyond
second-order.  In going beyond second-order, the induced electrostatic
interaction with the solvent is a nonlinear function of the solute
charges.  It is natural to think of these higher-order contributions
in terms of the concepts of field-strength-dependent dielectric
constant (saturation) and a field-strength-dependent local density
(electrostriction).  The molecular theory Eq.~(5) properly
incorporates effects that might be ascribed to a `local dielectric
constant' without addressing broader issues.  It will be important to
investigate higher order corrections to the simple theory of Eq.~(5).
In this context, Rashin [3] has suggested that contraction of cavity
radii about the incipient doubly charged ionic solute $Cl^- \cdots
Cl^-$, when that pair adopts solvent-excluding configurations, might
lead to a minimum in the potential of mean force.  Since a minimum is
observed in the molecular results for the $Cl^- \cdots Cl^-$ pair
potential of mean force {\it only} for solvent-separated
configurations (see Fig. 1), that argument is not considered further
here.

	Roux, {\it et al.\/}, [31] have presented an analysis starting with
the XRISM approximation which gives a different perspective on the
Born approximation for the solvation free energy of a spherical ion.
They helpfully focused particular attention on ions of the same
physical size but of opposite charge, {\it i.e.\/}, {\it Cl}$^-$ and
the hypothetical `{\it Cl}$^+$.' That previous analysis did not
identify the contribution linear in the charges that the much more
general discussion here fixes as the primal neglect of the dielectric
models.

	The calculation of the perturbative contributions of Eq.~(5) should
be quite feasible.  Such calculations would require simulation of
solutions without solute-solvent electrostatic forces for one solute
molecule.  The mean electrostatic solute-solvent interaction energy
and the mean square fluctuation of that quantity must be measured.  As
might be guessed by the correspondence with Eq.~(4) the fluctuation
term raises computational issues similar to those that arise in
calculation of the dielectric constant of the solvent, namely the
result can be sensitive to treatment of long-ranged interactions.  A
satisfactory calculation of this term will require further theoretical
consideration of those issues than we give here.  Where the dielectric
models are physically correct, the proposed calculation should provide
a accurate theory for the solvation free energy because it has a
similar physical origin as the dielectric models but augments those
models with the proper molecular details.

  	To the extent that the perturbative approach is valid, it can be
worthwhile to reexpress it in alternative ways which are physically
equivalent.  Indeed, Eq.~(5) deserves a fuller theoretical analysis to
be more useful for parameterization of dielectric models.  We expect
to pursue that theoretical analysis at a later date.

The simulation calculations of Jayaram, {\it et al.\/}, [32] designed
to test the Born model of solvation for spherical ions, cast some
light on the operation of these formulae.  Those calculations gave
clear evidence of structural saturation --- in the sense discussed by
Stell [36] --- by the time the charge on the ion was $q \approx e $.
See Fig.~(3) of Ref. (32) which showed that the change in solvation
free energy with respect to charge of the ion was approximately linear
for small and for large $q$ but the intercepts and slopes were
different in the two regimes. The cross-over between the two regimes
was near $q \approx e $.  This was just the region where separate
structural data began to show saturation; see Fig.~(4a) of that work.

	On the basis of those results we conclude that the perturbation
theory should be satisfactory for simple monovalent inorganic ions.
For polar molecules, charge concentrations which exceed $q \approx e$
per atom are unusual; hydrogen-bonding interactions are essentially
electrostatic and might offer exceptions to this hypothesis because
those charges can be well exposed.  Aside from such separate cases,
the perturbative approach is expected to be helpfully accurate for
solvation of polar solutes.  The description of the solvation of
multivalent inorganic ions on the basis of this perturbative approach
would be expected to incur more substantial inaccuracies.  Complex
ions of higher formal charge such as $CO_3^{2+}$ might also require
additional attention.  In both of the latter two cases the chemical
nature of the near-neighbor interactions is typically of paramount
importance and such interactions have not been considered at all here.

\bigskip \noindent {\bf Illustrative Example}

	We note again that the calculations called-for by Eq.~(5) treat a
solution with one solute for which solute-solvent electrostatic
interactions are extinguished.  It is possible to use a second-order
perturbation theory to calculation $\Delta \mu_0$ on the basis of
simulation results for the fully coupled solute and that inverse
approach is a likely to be a helpful technical convenience. The reason
for that convenience is that one utility of these theoretical ideas
would be the ability to calculate free energies on the basis of a
reliable physical theory without intrusive computational devices that
are typical of more general calculations of free energies, {\it
e.g.,\/} umbrellas and stratifiers.  The inverse formula will be
accurate to the extent that Eq~(5) is quantitatively accurate.  In
either the direct approach of Eq.~(5) or the inverse approach $\Delta
\mu_0$ must be separately provided.  To present an illustration here,
we have adopted the direct approach of Eq.~(5) and carried out short
molecular dynamics simulations (MD) on a series of systems consisting
of 215 water (TIP3P) molecules and one neutral van der Waals solute,
denoted by `M,' with diameters of 0.20, 0.30, 0.325, 0.350, 0.373,
0.40, and 0.45 nm. The van der Waals minimum energy depth for the
solute, $\epsilon$, was kept at 0.294 Kcal/mol. Solute-solute
interaction parameters are obtained from the relations $A^2_{MM}=
4\epsilon \sigma ^{12}$, $C^2_{MM} = 4\epsilon \sigma^6$, the
solute-solvent interactions from $A_{OM} = A_{OO} A_{MM}$, --- with
`O' indicating the solvent oxygen atom --- and $C_{OM} =
C_{OO}C_{MM}$, where $A_{OO}$ and $C_{OO}$ are defined by the TIP3P
model [37]. The systems were contained in a cubic box of width $L$ =
1.868 nm. Constant energy and volume MD simulations were done on each
system with average temperatures of 292 ($\pm$ 10K) and a water
density of 0.99 g/cm$^3$.  Electrostatic forces were evaluated with a
generalized reaction field (GRF) [38] and with Ewald summation (ES)
[18] approaches. For the GRF all interactions were truncated at $R_c
=\ 0.75$ nm, and the reaction field dielectric constant was taken to
be 65.0. For the Ewald sum, $2 \times 510$ {\bf k}-vectors were used
in the reciprocal space part of the potential, the screening constant
was 5.5/$L$, direct interactions were truncated at $L$/2, and the
reaction field dielectric constant was 65.0 again. The average
potential at the center of the van der Waals solute atom was
calculated using the GRF, the Ewald potential, and from integrals over
the obtained solute-water pair correlation functions using the bare
Coulomb and the GRF potentials.  Calculations with the GRF were
extended for 125 ps while those with the ES were extended for 50 ps.
In order to assess the size dependence of our results, one 50 ps
simulation of a system containing 511 waters and one solute atom
($\sigma$ =0.373 nm), using the ES, was done. The box size for this
system is 2.484 nm.  Unless otherwise indicated, the reported average
potentials were calculated with the GRF.

An important route for improvement of the dielectric models is to
obtain a separate assessment of the first-order perturbative term and
then establish good values for the cavity radii on the basis of the
second order term. Fig. (4) shows the dependence of the average
electrostatic energy of interaction of the solute and how it depends
on solute size. The indicated error bars are $\pm$ one estimated
standard deviation.  Within these uncertainties a trend with the
solute size cannot be confidently established. However, this
contribution is positive and substantial in magnitude. The fact that
the average potential at the center of the uncharged solute atom is
positive has been verified for both the Ewald and the GRF models.
These results can be understood in terms of the structure of water
near the solute.  Fig. (5) shows the distribution functions describing
the average radial disposition of the solvent charge about the solute
obtained with the Ewald (215 waters + solute) model.  The first peak
of the solute-hydrogen (M-H) rdf is broader than the peak for the
solute-oxygen peak, which is located at about the same distance.  As a
result there is a net positive charge density near the solute. Also
shown in Fig.(5) are: the charge density around the solute, defined by
$q(r) = 2q_H\bigl( g_{MH}(r) - g_{MO}(r) \bigr) $ and the integrated
charge density, $$ C(R) = 4 \pi \int_0^{R} \bigl( g_{MH}(r) -
g_{MO}(r) \bigr) r^2 dr \ , \eqno{(6)}$$ as it depends on the radius,
$R$, of included solvent charges from the spherical solute.  Fig. (6)
shows the contributions to the Coulomb and GRF potentials at the
center of the solute atom arising from solvent molecules. This
function is obtained by $$ V(R) = 4 \pi(2 q q_H) \int_0^R r^2
u(r)q(r)dr.  \eqno{(7)}$$ Here $u(r) = 1/r$ for the Coulomb potential
and $$ u(r) = u_{GRF} (r) = {1 \over r} ( 1 - {r\over R_c} )^4 (1 + {8
r \over 5R_c} + { 2 r^2 \over 5 R_c^2} )\Theta(R_c -r) \eqno{(8)} $$
with $\Theta(R_c -r)$ the Heaviside step function, for the GRF.
Notice that the GRF rapidly converges to the average value, while
contributions to the bare Coulomb potential, $1/r$, oscillate with
increasing $R$ around the average value even at large distance from
the solute.  As a result we believe that averages over the bare
Coulomb potential in finite size boxes will be sensitive to the system
size and simulation parameters.

 The average value of the potential obtained by the Ewald sum for two
system sizes (215 and 511 water molecules) are shown by the cross and
triangle labels in Fig. (6).  A positive average potential energy
implies that the solvation of a negative ion is lower (more favorable)
than the solvation of a positive ion of the same physical size.  The
radii recommended by Rashin and Honig [21] for the isoelectronic $K^+$
and $Cl^-$ ions are 0.2172 and 0.1937 nm, respectively. Within the
Born picture, the physical size of the $K^+$ ion should be slightly
smaller than that of the Cl$^-$ ion. Thus, these recommended empirical
radii are consistent with our calculation of a {\it positive\/}
potential at zero charge.  This relative ordering of recommended radii
is observed for all isoelectronic (alkali metal ion, halide ion) pairs
[21].  The present explanation of the ordering of empirical radii for
isoelectronic ions is quite different from one of Rashin and Honig
[21] which invokes quantum effects and suggests ``the electron cloud
of the anion is unable to significantly penetrate the empty valence
orbital of the cation.'' [21] The present explanation is consistant
with but distinct from the intuitive explanation of Latimer, {\it et
al.\/}, [39] that anion radii are smaller than isoelectronic cation
radii ``since the negative ions will have the hydrogens directed
in,$\ldots$'' The present explanation uses quantitative information on
the disposition of the water molecules observed about a modelled {\it
neutral\/} atom of the same physical size.  The inference of a {\it
positive\/} potential at zero charge on the basis of the observed
ordering of empirical radii of isoelectronic alkali and halide ions
appears to be new.

 The evaluation of the second term in the perturbation scheme in Eq.
(5) is expected to be sensitive to the modeling of the solvent, both
to the assumed molecular structures and forces, and to the treatment
of long-ranged (electrostatic) interactions.  An important point to be
considered is the contribution of electrostatic self-energy terms to
the total energy of the system. In adding a charged particle to the
system, a constant self-energy contribution proportional to $q^2$ is
added to the energy differences between the charged and uncharged
systems. However, different schemes for modeling the long-ranged
electrostatic interactions imply different self-energy contributions,
even though the total energies are the same.  It is natural to group
these terms together with the fluctuation, $ \langle( \sum_j
\varphi(j) - \langle \sum_k \varphi(k) \rangle_0 )^2 \rangle_0 $,
because they are both proportional to $q^2$.  As a result, it is
expected that the variances in the potential will differ from one
approach to another, but when the self-energy terms are correctly
included, all models should give similar results for terms
proportional to $q^2$.  These points deserve more specific theoretical
study and will be discussed elsewhere.  Fig.~(7) shows the dependence
of the mean square fluctuations of the potential energy as a function
of the size of the solute in our calculations --- without the
self-energy terms in the GRF model.

Finally, we inquire more specifically about the accuracy of the
perturbative approach. In principal, we could calculate succeeding
terms in the perturbation series. In practice, such an approach will
not avail us here. However, the cumulant expansion to second order in
$\beta $ $$ \exp[ -\beta (\Delta \mu - \Delta \mu_0) ] = \langle \exp(
-\beta [U(q=1) - U(q=0)] \rangle_0 $$ $$ \sim \exp[ -\beta \langle
\sum_j \varphi(j)\rangle_0 + {\beta^2 \over 2} \langle( \sum_j
\varphi(j) - \langle \sum_k \varphi(k) \rangle_0 )^2 \rangle_0 ],
\eqno{(9)} $$ is exact for a Gaussian distribution of the variable
$\sum_j \varphi(j)$.  We can verify that the perturbation approach is
sensible by noting the validity of a Gaussian distribution of the
variable $\sum_j \varphi(j) $ when observed in the system for which no
solute-solvent electrostatic interactions are expressed. Fig. (8)
shows distributions of that variable observed for large (0.373 nm) and
small (0.20 nm) solutes considered and, in addition, gaussian model
distributions fitted to the observations. The detailed correctness of
the perturbation theory will be sensitive to the Gaussian character of
these distributions {\it in the wings\/} of the distributions and
those characteristics are not tested very critically here.  However,
the agreement with the Gaussian model is satisfactory in the sense
that any differences are unlikely to be statistically significant on
the basis of the data.

\bigskip \noindent {\bf Conclusions}

This work has confirmed that the surprising qualitative features of
the Rashin's dielectric model calculations of alkali-halide
potentials-of-mean-force in water are correct results of the model.
However, the molecular information to which he compared his results
were not of sufficient accuracy to conclude that the dielectric model
gave a valid description of the free energies of inter-ionic
interactions on a thermal scale.  Thus, the dielectric model results,
although surprising and physical, are not accurate descriptions of the
inter-ionic potentials-of-mean-force for alkali-halide/water solution
as those quantities are presently known.  Other results suggest that a
dielectric model can be helpfully accurate on a chemical energy scale.
But the accuracy of free energies to thermal levels, {\it i.e.\/}, to
better than $kT$ for molecular scale rearrangments, has not been
broadly tested.  It must be emphasized that such tests have been
limited by the lack of accurate molecular results as standards for
comparison.  One exception is the work of Wilson and Pohorille [34]
which studied the interaction of monovalent ions with the water
liquid-vapor interface.  Because of the structural subtlety of those
problems they are likely to remain difficult challenges for physical,
molecular thermodynamic models.

The extant favorable predictions of dielectric models highlight the
most crucial limitation for their application: the definition of
molecular volumes or cavity radii.  The dielectric model is precisely
a modelistic implementation of second-order perturbation theory for
the excess chemical potential of the solute. Thus the straightforward
theoretical procedure for the study of how that molecular volume
should be defined can be based upon thermodynamic perturbation theory
evaluated through second-order.  The first-order term in that
perturbation theory, or equivalently the {\it potential at zero
charge}, vanishes in the dielectric models but is generally nonzero.
Thus an assessment of that first-order contribution is the most
important step for further development of dielectric models --- of
more immediate concern than higher-order perturbative contributions.
The {\it positive\/} value of this potential at zero charge computed
for atomic solutes in water gives a very basic explanation of the
observation that empirical radii for halide ions are slightly smaller
than the empirical radii for isoelectronic alkali metal ions.  With
knowledge of this first-order term, parameterization of radii and
molecular volumes should then be based of the second-order
perturbative term alone.

The second-order perturbative theory can be expressed compactly and
should be quite feasible to compute.  This fluctuation term raises
computational issues similar to those that arises in calculation of
the dielectric constant of the solvent regarding treatment of
long-ranged interactions.  Satisfactory calculation of that term will
require additional theoretical consideration of those issues.  To the
extent that the dielectric approach is reasonable, second-order
thermodynamic perturbation theory for the excess chemical potential of
the solute, implemented on a molecular basis, should provide a simple
and accurate theory of solvation thermodynamics associated with
electrostatic interactions.

\bigskip \noindent {\bf Acknowledgements}

One of us (LRP) thanks Andrew Pohorille and Greg Tawa for helpful
discussions and also gratefully acknowledges the support for this work
in part from the Tank Waste Remediation System (TWRS) Technology
Application program, under the sponsorship of the U. S. Department of
Energy EM-36, Hanford Program Office, and the Air Force Civil
Engineering Support Agency.  This work was also supported in part by
the US-DOE under Los Alamos National Laboratory LDRD-PD research
funds.

\vfil \eject

\beginsection{References}

\item{1.} H. L. Friedman and W. D. T. Dale, in MODERN THEORETICAL
CHEMISTRY, Vol. 5, edited by B. Berne (Plenum, NY, 1977).  Chapter 3,
``Electrolyte solutions at equilibrium''

\item{2.} J. P. Valleau and G. M. Torrie, in MODERN THEORETICAL
CHEMISTRY, Vol. 5, edited by B. Berne (Plenum, NY, 1977).  Chapter 5,
``A guide to Monte Carlo for statistical mechanics: 2.  Byways.'' See
particularly Section 4.2.

\item{3.} A. A. Rashin, {\it J. Phys. Chem.\/} {\bf 93} (1989)4664.

\item{4.} D. Chandler, in THE LIQUID STATE OF MATTER: FLUIDS, SIMPLE
AND COMPLEX, edited by E. W. Montroll and J. L. Lebowitz
(North-Holland, Amsterdam, 1982). p. 275.

\item{5.} P. T. Cummings and G. Stell, {\it Mol. Phys.\/} {\bf 46}
(1982)383.

\item{6.} P. A. Monson and G. P. Morriss, in ADVANCES IN CHEMICAL
PHYS\-ICS, Vol. 77, edited by I. Prigogine and S. A. Rice (Wiley, NY,
1990), p. 451.

\item{7.} D. Chandler, R. Silbey, and B. M. Ladanyi, {\it Mol.
Phys.\/} {\bf 46} (1982)1335.

\item{8.} D. Chandler, J. D. McCoy, and S. J. Singer, {\it J. Chem.
Phys.\/} {\bf 85} (1986)5977.

\item{9.} G. Hummer and D. M. Soumpasis, {\it Mol. Phys.\/} {\bf 75}
(1992)633.

\item{10.} (a) J. S. Perkyns and B. M. Pettitt, {\it Chem. Phys.
Lett.\/} {\bf 190} (1992)626; (b) J. Perkyns and B. M. Pettitt, {\it
J. Chem. Phys.\/} {\bf 97} (1992)7656.

\item{11.} P. J. Rossky, {\it Ann. Rev. Phys. Chem.\/} {\bf 36}
(1985)321.

\item{12.} B. M. Pettitt and P. J. Rossky, {\it J. Chem. Phys.\/} {\bf
84} (1986)5836.

\item{13.} H. L. Friedman, F. O. Raineri, and H. Xu, {\it Pure Appl.
Chem.\/} {\bf 63} (1991)1347.

\item{14.} (a) G. Hummer, D. M. Soumpasis, and M. Neumann,
{\it Mol. Phys.\/} {\bf 77} (1992)769;
(b) {\it Mol. Phys.\/} {\bf 81} (1994)1155.

\item{15.} J. S. H{\o}ye and G. Stell, {\it J. Chem. Phys.\/} {\bf 65}
(1976)18.

\item{16.} H. J. C. Berendsen, J. P. M. Postma, W. F. van Gunsteren,
and J. Hermans, in JERUSALEM SYMPOSIA ON QUANTUM CHEMISTRY AND
BIOCHEMISTRY, B. Pullman, editor (Reidel, Dordrecht, 1981), p. 331.

\item{17.} L. X. Dang, B. M. Pettit, and P. J. Rossky, {\it J. Chem.
Phys.\/} {\bf 96} (1992)4046.

\item{18.} M. P. Allen and D. J. Tildesley, COMPUTER SIMULATION OF
LIQUIDS (Clarendon, Oxford, 1987).

\item{19.} E. Gu\`{a}rdia, R. Rey, and J. A. Padr\'{o}, {\it Chem.
Phys.\/} {\bf 155} (1991)187.

\item{20.} G. Hummer, D. M. Soumpasis, and M. Neumann, {\it J. Phys.
Condens. Matter\/} (in press, 1994).

\item{21.} A. A. Rashin and B. Honig, {\it J. Phys. Chem.\/} {\bf 89}
(1985)5588.

\item{22.} T. J. You and S. C. Harvey, {\it J. Comp. Chem.\/} {\bf 14}
(1983)484.

\item{23.} J. L. Pascual-Ahuir, E. Silla, J. Tomasi, R. Bonaccorsi,
{\it J. Comp. Chem.\/} {\bf 8} (1987)778.

\item{24.} R. J. Zauhar and R. S. Morgan, {\it J. Mol. Biol.\/} {\bf
186} 815(1985); {\it J. Comp. Chem.\/} {\bf 9} 171(1988).

\item{25.} A. A. Rashin and K. Namboodiri, {\it J. Phys. Chem.\/},
{\bf 91} (1987)6003; {\it J. Phys. Chem.\/} {\bf 94} (1990)1725.

\item{26.} B. J. Yoon and A. M. Lenhoff, {\it J. Comp. Chem.\/} {\bf
11} (1990)1080; {\it J. Phys. Chem.\/} {\bf 96} (1992)3130.

\item{27.} A. H. Juffer, E. F. F. Botta, B. A. M. van Keulen, A. van
der Ploeg, and H. J. C. Berendsen, {\it J. Comp. Phys.\/} {\bf 97}
(1991)144.

\item{28.} B. Wang and G. P. Ford, {\it J. Chem. Phys.\/} {\bf 97}
(1992)4162.

\item{29.} B. Honig, K. Sharp, and A.-S. Yang, {\it J. Phys. Chem.\/}
{\bf 97} (1993)1101.

 \item{30.} A. Jean-Charles, A. Nicholls, K. Sharp, B. Honig, A.
Tempczyk, T. F. Hendrickson, and W. C. Still, {\it J. Am. Chem.
Soc.\/} {\bf 113} (1991)1454.

\item{31.} B. Roux, H.-A. Yu, and M. Karplus, {\it J. Phys. Chem.\/}
{\bf 94} 4683(1990).

\item{32.} B. Jayaram, R. Fine, K. Sharp, and B. Honig, {\it J. Phys.
Chem.\/} {\bf 93} (1989)4320.

\item{33.} R. M. Levy, M. Belhadj, and D. B. Kitchen {\it J. Chem.
Phys.\/} {\bf 95} (1991) 3627.

\item{34.} L. R. Pratt, {\it J. Phys. Chem.\/} {\bf 96} (1991)25.

\item{35.} M. A. Wilson and A. Pohorille, {\it J. Chem. Phys.\/} {\bf
95} (1991)6005.

\item{36.} G. Stell, in MODERN THEORETICAL CHEMISTRY, Vol. 5, edited
by B. Berne (Plenum, NY, 1977).  Chapter 2, ``Fluids with long-range
forces: towards a simple analytic theory''

\item{37.} W. L. Jorgensen, {\it J. Am. Chem. Soc.} {\bf 103}
(1981)339; see also W.L. Jorgensen, J. Chandrasekher, and J.D. Madura,
{\it J. Chem. Phys.\/} {\bf 79} (1983)926.

\item{38.} G. Hummer and D.M. Soumpasis, {\it Phys. Rev. E}, (in
press, 1994).

\item{39.} W. M. Latimer, K. S. Pitzer, and C. M. Slansky, {\it J.
Chem. Phys.\/} {\bf 7} (1939)108.

\vfil \eject

\beginsection{Figure Captions}

\item{Fig. (1):} Potentials-of-mean-force for (top panel) $Na^+ \cdots
Na^+$, (middle panel) $Na^+ \cdots Cl^-$, and (lower panel) $Cl^-
\cdots Cl^-$ in water as determined from the molecular dynamics
calculations of Ref.~14.  The solid line in each panel is the result
for the room temperature system and the dashed line is the result for
the system at $T=823$~K.

\item{Fig. (2):} Potential-of-mean-force for the $Na^+ \cdots Cl^-$
ion pair predicted by a dielectric model for water at its triple
point.

\item{Fig. (3):} Separate contributions to the potential-of-mean-force
of Fig.~(2).  The solvation free energy of the ion pair is $\Delta \mu
(r)$ and the bare inter-ionic interaction potential energy is $u(r)$.

\item{Fig. (4):} The average electrostatic potential of the solvent
water at the center of spherical van der Waals solute as a function of
size of the solute.  The error bars indicating one standard deviation
were estimated by block averaging over 10 ps segments of the
trajectories.

\item{Fig. (5):} Radial distribution of solvent (water) atoms about
from the center of a spherical van der Waals solute with diameter 0.2
nm.  The solid curve is the solute-oxygen radial distribution function
$g_{MO}(r)$; the dashed curve is the solute-hydrogen radial
distribution function $g_{MH}(r)$; the dash-dot curve shows the
integrated charge, $C(r)$ of Eq.~(6), enclosed in a ball of the
specified radius; the dash-dot-dot curve plots the difference function
$g_{MH}(r)-g_{MO}(r)$.

\item{Fig. (6):} Electrostatic potential in energy units at the center
of a neutral solute atom showing dependence on methods of calculation
and system size.  The solute is a Lennard-Jones atom with $\sigma = $
0.373 nm as discussed in the text.  The designation `Coulomb' (heavy
lines) or `GRF' (light lines) indicate how the electrostatic potential
at the solute was calculated according to Eqs.~(7) and (8).  In
information is parentheses gives the number of water molecules in
system studied, either 511 or 215, and the method of treatment of the
long-ranged electrostatic interactions between water molecules, either
Ewald summation (ES) or generalized reaction field (GRF).  The cross
(plotted at $r = $ 0.9 nm) and triangle (at $r = $ 1.2 nm) are the
average Ewald potentials (12.8 $\pm$ 1.8 and 9.8 $\pm$ 1.7 Kcal/mol
for the N=215 and N=511 systems, respectively) obtained during the
simulations. The corresponding values obtained with GRF are plotted in
Fig. (4).

\item{Fig. (7):} Mean square fluctuations of the potential at the
center of the van der Waals solute as a function of VdW diameters for
a set of simulations with N=215 waters using the GRF approach.

\item{Fig. (8):} Distributions of the observed solvent electrostatic
potential $ \sum_j \varphi({\bf j}) $ at the center of the van der
Waals solute for different size solutes considered.  The dashed curves
are gaussian distributions fitted to the observed data.

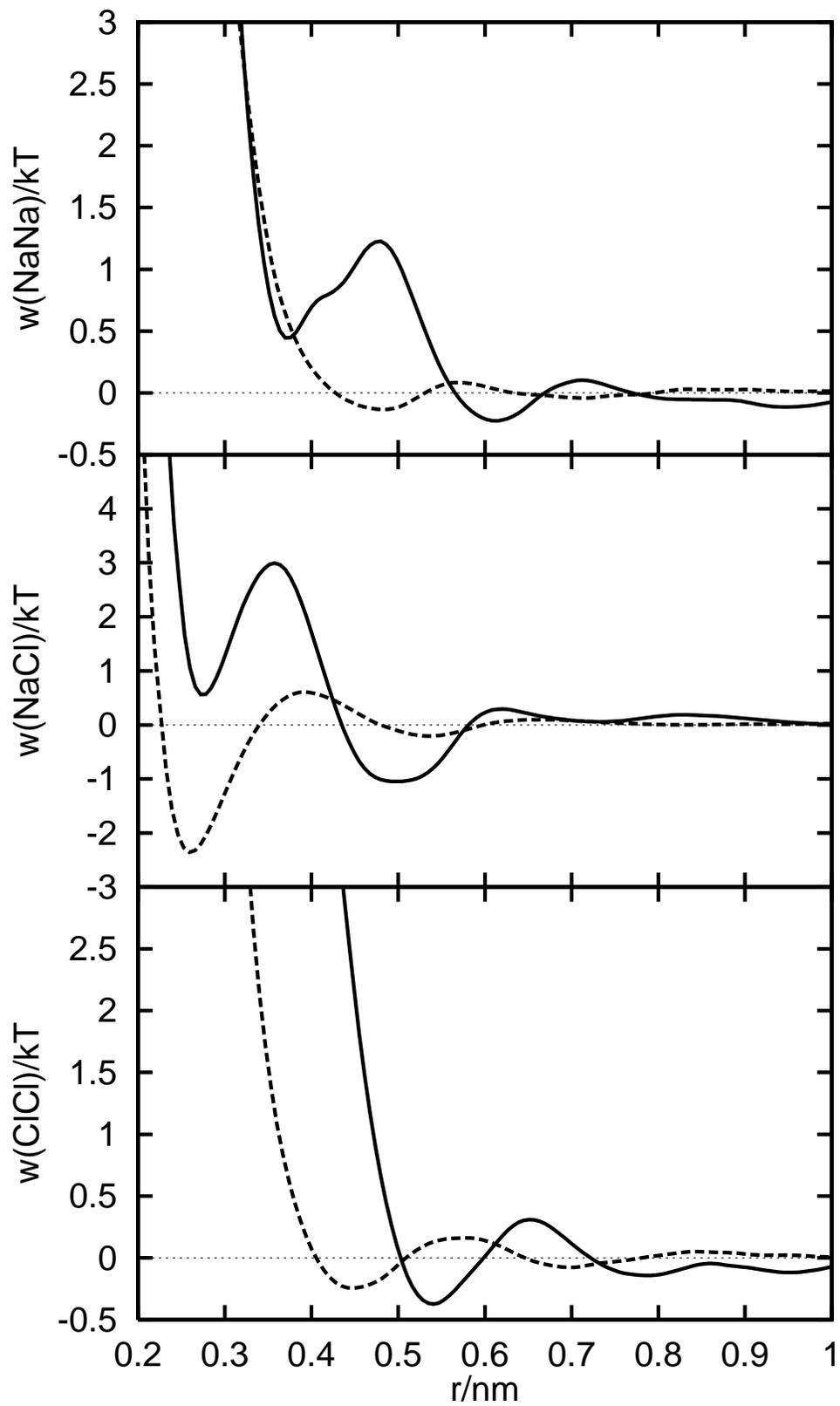
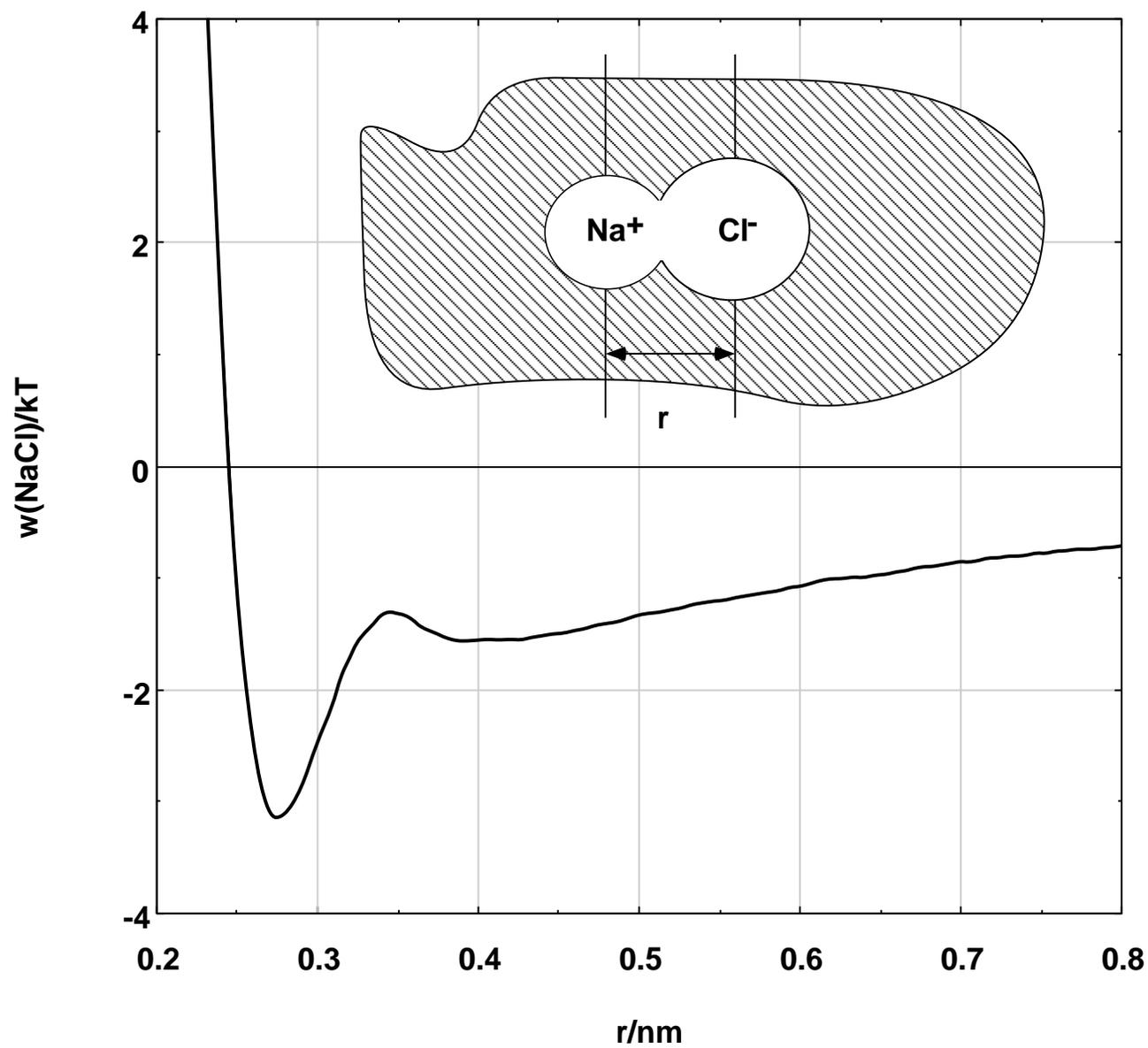
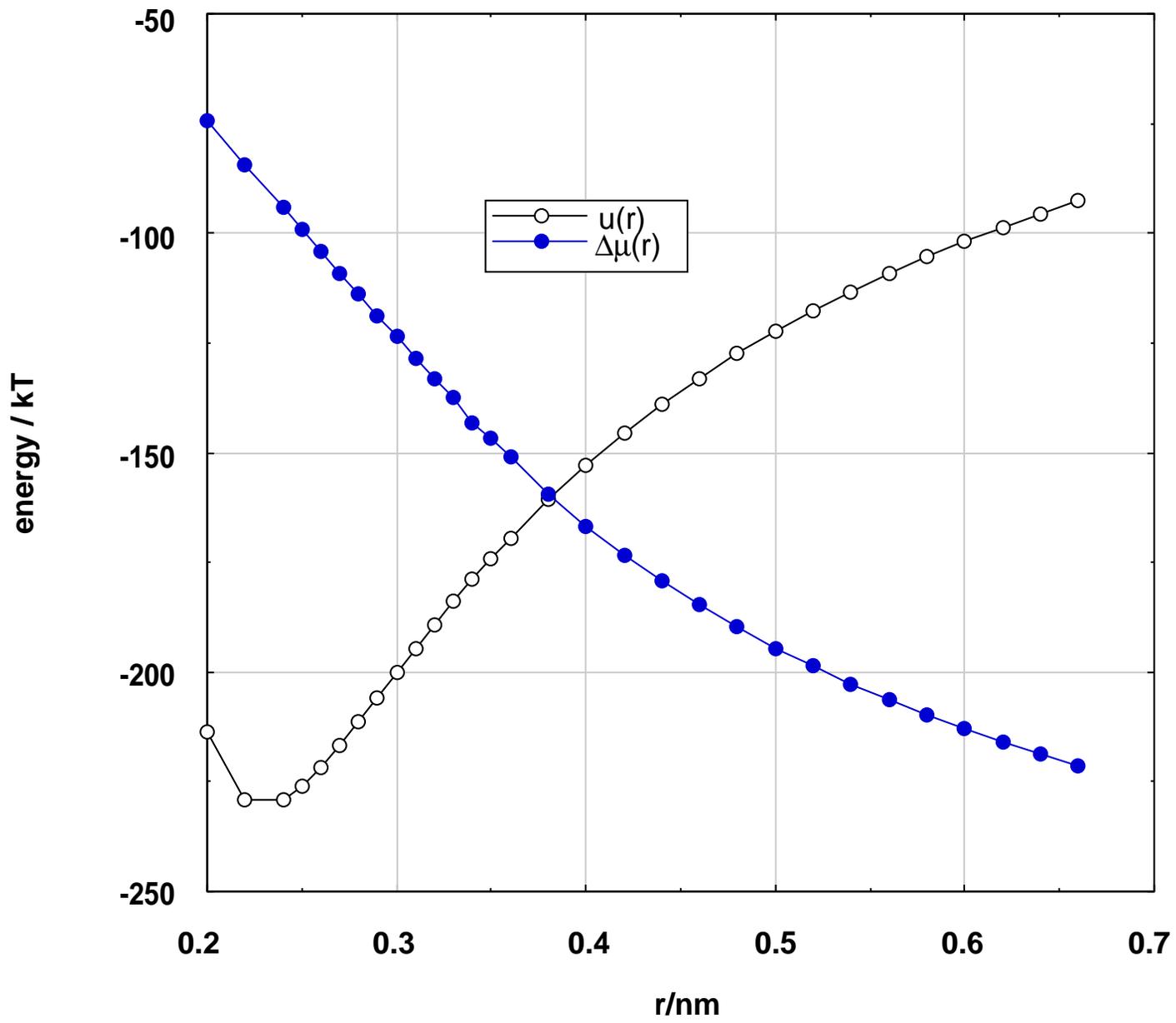
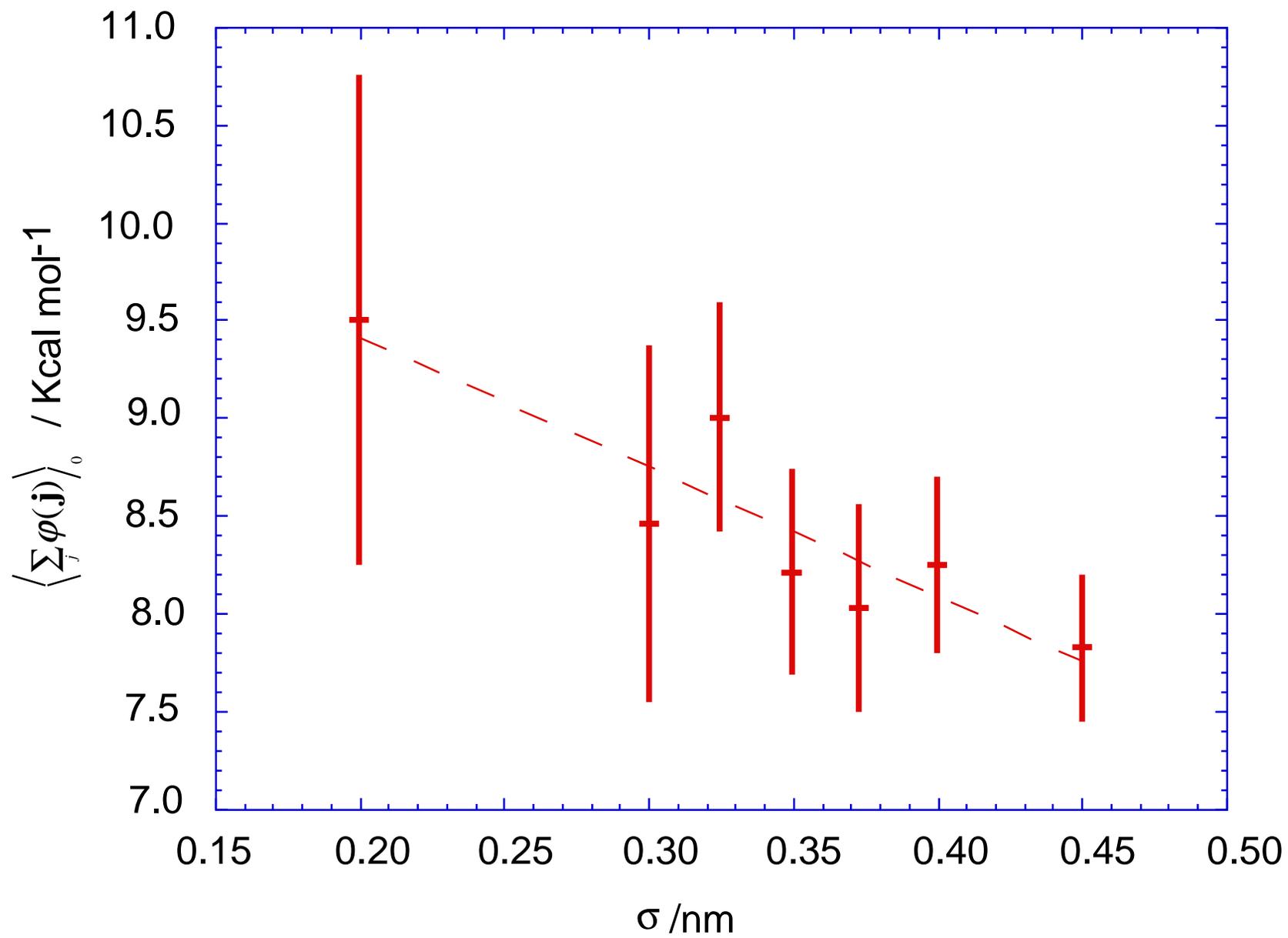
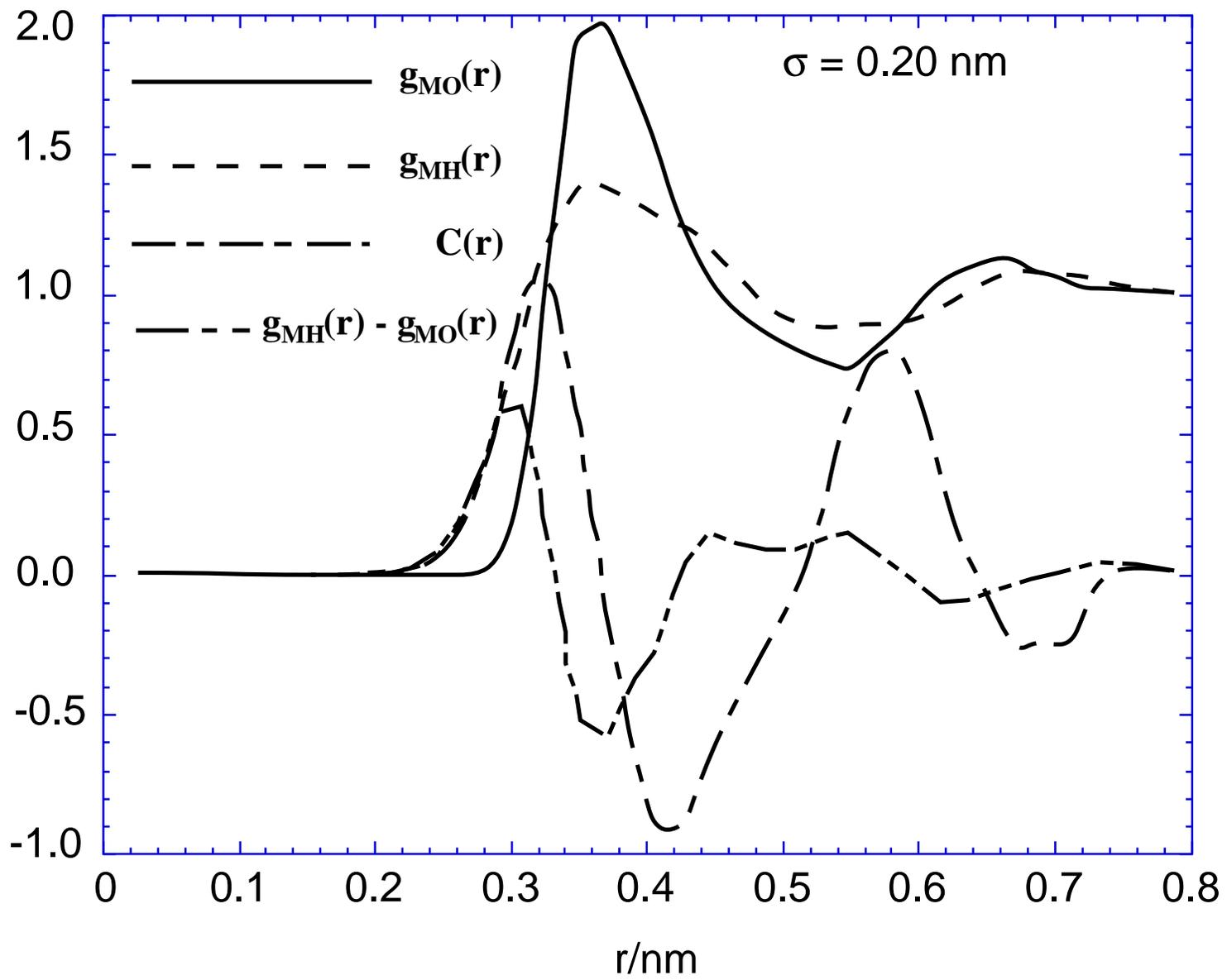
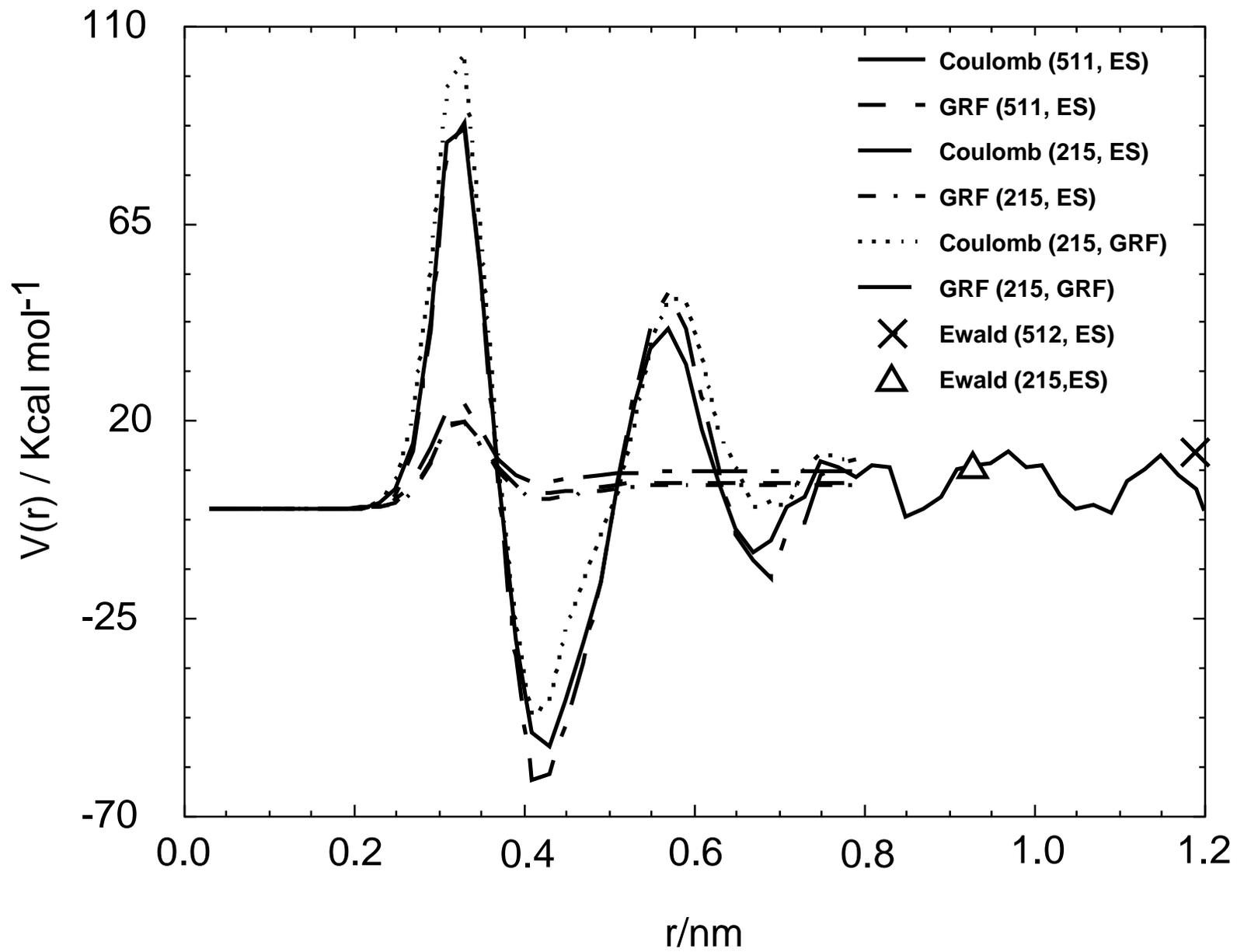
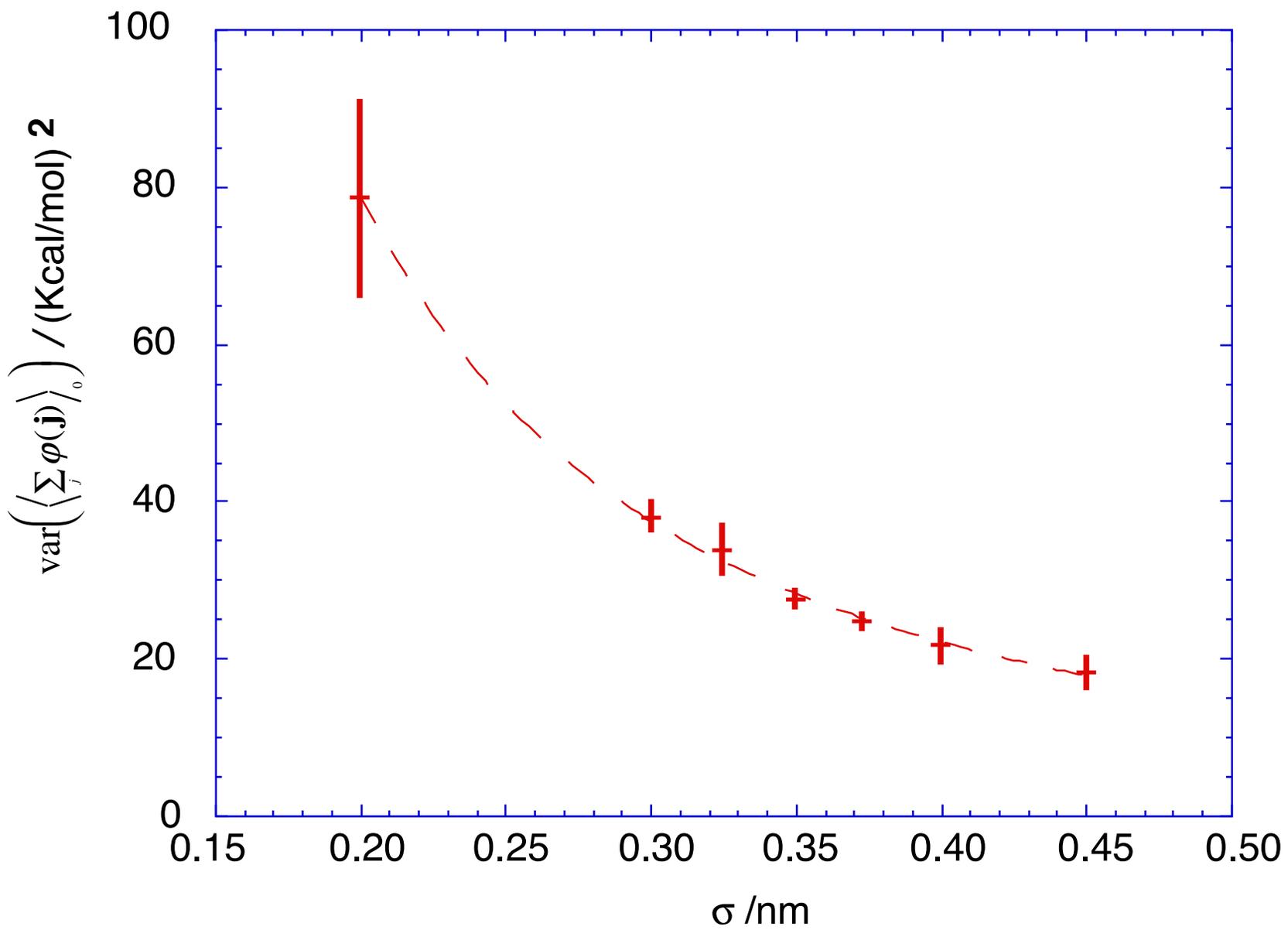
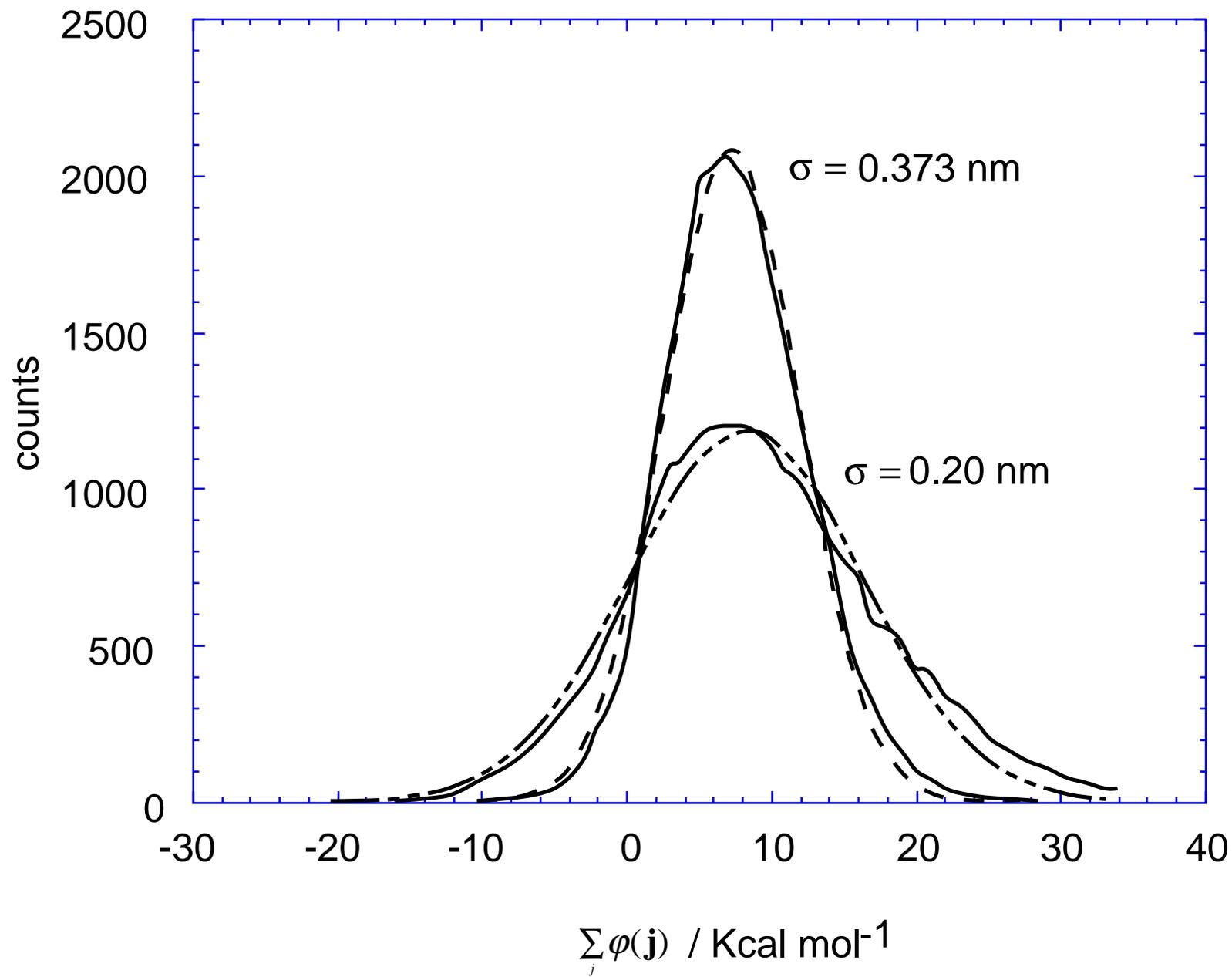
\end